\documentclass[final,english]{bullsrsl}[2022/06/15]

\usepackage[latin1]{inputenc}
\usepackage[T1]{fontenc}

\usepackage{natbib} 
\usepackage{graphicx}

\begin{document}
\title{Insights on the Optical and Infrared Nature of MAXI J0709-159: Implications for High-Mass X-ray Binaries}

\author[affil={1}, corresponding]{Suman}{Bhattacharyya}
\author[affil={1}]{Blesson}{Mathew}
\author[affil={1}]{Gourav}{Banerjee}
\author[affil={2}]{Sindhu}{G}
\author[affil={3}]{S.}{Muneer}
\author[affil={3}]{S.}{Pramod Kumar}
\author[affil={4}]{Santosh}{Joshi}

\affiliation[1]{Department of Physics and Electronics, CHRIST (Deemed to be University), Hosur Main Road, Bangalore, India}
\affiliation[2]{Department of Physics, St. Thomas College, Kozhencherry, Kerala-689641, India}
\affiliation[3]{Indian Institute of Astrophysics, Koramangala, Bangalore, India}
\affiliation[4]{Aryabhatta Research Institute of Observational Sciences (ARIES), Nainital, India}

\correspondance{suman.bhattacharyya@res.christuniversity.in}



\maketitle


%

\begin{abstract}

In our previous study \citep{2022Bhattacharyya}, HD~54786, the optical counterpart of the MAXI J0709-159 system, was identified to be an evolved star, departing from the main sequence, based on comparisons with non-X-ray binary systems. In this paper, using color-magnitude diagram (CMD) analysis for High-Mass X-ray Binaries (HMXBs) and statistical t-tests, we found evidence supporting HD 54786's potential membership in both Be/X-ray binaries (BeXRBs) and supergaint X-ray binaries (SgXBs) populations of HMXBs. Hence, our study points towards dual optical characteristics of HD~54786, as an X-ray binary star and also belonging to a distinct evolutionary phase from BeXRB towards SgXB. Our further analysis suggests that MAXI J0709-159, associated with HD 54786, exhibits low-level activity during the current epoch and possesses a limited amount of circumstellar material. Although similarities with the previously studied BeXRB system LSI +61$^{\circ}$ 235 \citep{1994Coe} are noted, continued monitoring and data collection are essential to fully comprehend the complexities of MAXI J0709-159 and its evolutionary trajectory within the realm of HMXBs.


\end{abstract}


\section{Introduction}

High-mass X-ray binaries (HMXBs) are binary systems composed of a compact object (possibly a neutron star or a black hole) and a massive companion star, which can be typically an O or B type star \citep{2011Reig}. They are important sources of X-ray emission in the Galaxy and other star-forming regions, as well as probes of stellar evolution and feedback. HMXBs can be classified into different subtypes based on the properties of their companion stars and their X-ray variability. Two of the most prominent subtypes are Be/X-ray binaries (BeXRBs) and supergaint X-ray binaries (SgXBs) \citep{2011Reig}.

Be/X-ray binaries (BeXRBs) consist of a compact object orbiting a Be star, which is a non-supergaint B-type massive star that exhibits Balmer emission lines and infrared excess due to the presence of a circumstellar gaseous disc \citep[e.g.][]{2011Reig, 1982Rappaport}. The compact object accretes matter from the decretion disc \citep{2001OkazakiAndNegueruela} of the Be star. A study by \cite{2009Belczynski} found that neutron stars (NSs) are more frequently observed companion for BeXRBs. Many BeXRBs are transient sources that exhibit periodic or sporadic outbursts of X-ray emission, lasting from minutes to weeks, when the compact object interacts with the disc or the stellar wind originating from the Be star \citep[e.g.][]{2017Monageng, 2001OkazakiAndNegueruela}. The orbital periods of BeXRBs range from tens to hundreds of days, and the eccentricities are usually high. They are the most common subtype of HMXBs in the Galaxy and in nearby galaxies such as the Magellanic Clouds \citep{2011Reig}.

In a recent study, we performed a follow-up study \citep{2022Bhattacharyya} on the recent detection of two X-ray flaring events by MAXI/GSC observations in soft and hard X-rays from MAXI J0709-159 in the direction of HD 54786 (LY CMa), on 2022 January 25. Using optical spectroscopy and multi-epoch photometry, their study primarily focused on the nature of the optical counterpart of MAXI J0709-159, which is the less-studied Be star HD 54786. The authors estimated the star's effective temperature to be 20000 K and found that it is evolving off the main sequence in the Color-Magnitude Diagram. Their analysis also suggested that HD 54786 is a BeXRB system having a compact object companion, probably a neutron star. However, this study was based on a sample of catalogs of non X-ray binary systems. So we became motivated to perform further analysis of this object using some well known HMXB catalogs to complement the previous study and better understand such systems. In the present paper, we present an optical photometric analysis of the location of HD 54786 in relation to well known HMXB systems. By focusing on the optical properties of this system, we gain important insights about the circumstellar disc, which not only aids in better understanding the characteristics of HMXBs but also sheds light on the nature of standalone Be stars.

\section{Data}
For the current study, we utilized the data adopted from some important HMXB catalogs, such as those of \cite{2006liu}, \cite{2019Kretschmar} and \cite{2023Neumann}. We selected and extracted two major classes of HMXBs, supergaints and Be stars, as indicated by the classification flags provided in these catalogs. By adhering to this classification criteria, we categorized the flagged objects accordingly. HMXBs associated with Be stars are classified as the "Bexrb" category, representing BeXRBs. Likewise, any HMXB flagged or associated with a supergaint was included in the category labeled "Sgxrb", denoting Supergaint X-ray Binaries (SgXBs). SgXBs consist of a compact object orbiting a highly evolved, luminous supergaint star \citep{2011Reig}. These persistent sources display varying X-ray luminosity levels depending on the mass accretion rate from the supergaint's strong stellar wind. Having orbital periods spanning between days to months, SgXBs constitute approximately 30\% of the Galactic population of HMXBs \cite{drave2013}.

We identified a total of 65 systems associated with Be stars, which we classified as "Bexrb", and 33 systems associated with supergaints, classified as "Sgxrb". From this sample, we selected 37 "Bexrb" systems and 17 "Sgxrb" systems for further analysis since these stars have Gaia magnitudes (G, $G_{BP}$, and $G_{RP}$) data \citep{Gaia2021A&A...649A...1G}, Gaia photometric distance \citep{2021Bailerjonesdr3} and extinction parameters (obtained from \citealt{2019Green}) available.

\section{Results}
\subsection{Gaia CMD analysis}
In our previous study, we determined the photometric position of HD 54786 in the \textit{Gaia} \citep{Gaia2021A&A...649A...1G} color-magnitude diagram (CMD) and compared its location with samples of previously studied B-type \citep{2010Huang}, Be stars \citep[and references therein]{2021Bhattacharyya}, gaint stars \citep{Hohle2010AN....331..349H} and supergaints \citep{2021Georgy}. It was noticed that this star is located near to the top of the distribution of B and Be stars, situated inside the distribution of gaint stars and below that of supergaints (see Fig.\,2a in \citealp{2022Bhattacharyya}).


\begin{figure}[!h]
\centering
\includegraphics[width=0.75\textwidth]{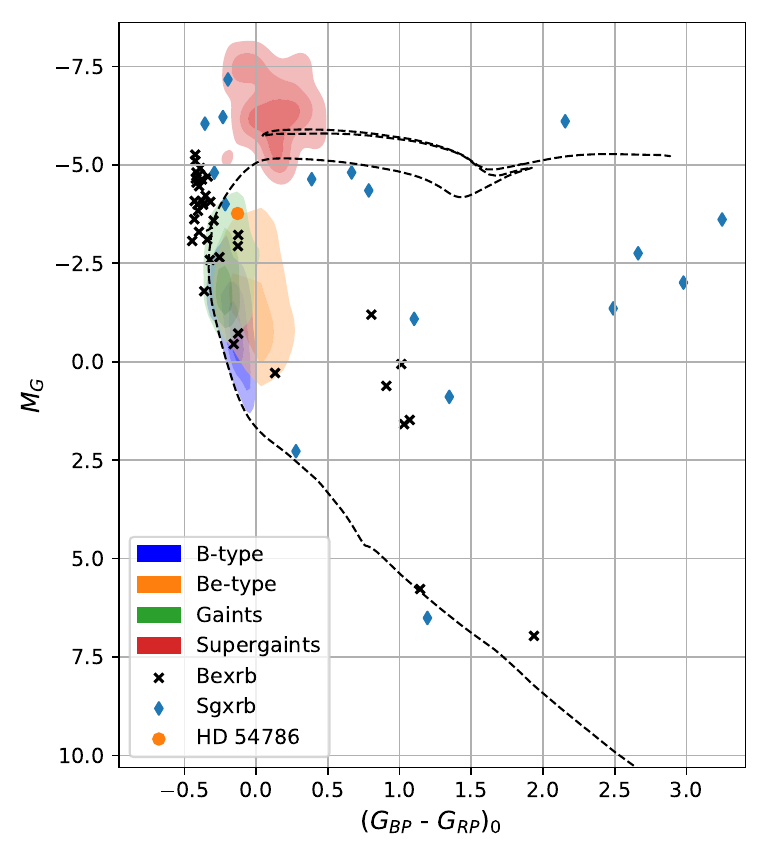}
\begin{minipage}{12cm}
\caption{The \textit{Gaia} CMD of HD 54786 having extinction corrected \textit{Gaia} M$_G$ and ($G_{BP}$ - $G_{RP})$$_0$ magnitudes has been represented. The \textit{Gaia} G and ($G_{BP}$ - $G_{RP})$ magnitudes are obtained from \protect \cite{Gaia2021A&A...649A...1G}. The probability distribution (Gaussian fitted at three contour levels) of the B, Be stars, gaints and supergaint are shown in blue, orange, green and red shaded colors, respectively \citep{2022Bhattacharyya}. The black dashed line in the plot represents the isochrone of 60 Myr with V/Vcrit = 0.4 (critical rotation fraction) and [Fe/H] = 0 (The top black dashed line is the blue loop part of the same isochrone). The black cross and the blue diamond markers indicates the "Bexrb" and "Sgxrb" systems, respectively.}
    \label{figure1}
    \end{minipage} 
\end{figure}

In this study, we constructed the \textit{Gaia} color-magnitude diagram (CMD) for HD 54786 using data from well known HMXB catalogs mentioned in Sect.\,2. Considering its \textit{Gaia} DR3 distance and A\textsubscript{V} value of 0.93 (as estimated from the Green's map; \citealp{2019Green}), we plotted the corresponding $M_G$ versus $G_{(BP-RP)0}$ CMD for the star and over-plotted the previously well-studied HMXB sample. It is noted from \cite{2013pecautmamajek} that the ZAMS line does not extend beyond B9 spectral type in $M_G$ and $G_{(BP-RP)0}$ values. So we utilised the closely matching 60 Myr isochrone track from MESA (Modules for Experiments in Stellar Astrophysics) isochrones and evolutionary tracks (MIST) \citep{2016choi, 2016dotter}, which is over-plotted in the CMD with V/V$_{crit}$ = 0.4 (critical rotation fraction), since that is the only model available in the MIST database for a rotating system as our sample shows features of Be star \citep{2022Bhattacharyya}. Moreover, we adopted the metallicity value of [Fe/H] = 0,  corresponding to solar metallicity $Z_\odot$ = 0.0142 \citep{2009Asplund} for the tracks. The CMD showing the location of HD 54786, with the representative locations of the selected extinction corrected "Bexrb" and "Sgxrb" systems is presented in Fig.\,\ref{figure1}.


The CMD shows distinct regions for the distributions of BeXRB and SgXB systems. In light of this, we sought to evaluate the involvement of HD 54786 in these two subclasses of HMXBs, i.e. "Bexrb" and "Sgxrb" distributions. To accomplish this, we conducted single t-tests using the Scipy Python module \citep{2020SciPy}. One-sample t-tests were chosen due to their ease of interpretation; if the obtained p-value is higher than a chosen significance level (e.g., 0.05), we can conclude that there is evidence of a significant similarity between the samples. In this study, the resulting p-values from the t-test for both the "Bexrb" and "Sgxrb" cases were higher than 0.05, suggesting that the results lacked statistical significance, failing to support the null hypothesis that HD 54786 does not belong to any distribution. The average p-values obtained were 0.3 and 0.4, respectively. Despite the higher p-value for the "Sgxrb" category, it is premature to classify HD 54786 as a supergaint associated system, given its proximity to the BeXRB value. Nonetheless, this analysis does support HD 54786's membership in both the "Bexrb" and "Sgxrb" populations, indicating its optical characteristics resemble those of X-ray binary stars, representing a distinctive stage in its evolutionary pathway towards SgXB from BeXRB.

This finding complements our previous results by specifically revealing the dual nature of HD 54786, thus further enriching our understanding of its evolutionary significance.


\subsection{H$\alpha$ vs J-K Study} 

\begin{figure}[t]
\centering
\includegraphics[width=1.0\textwidth]{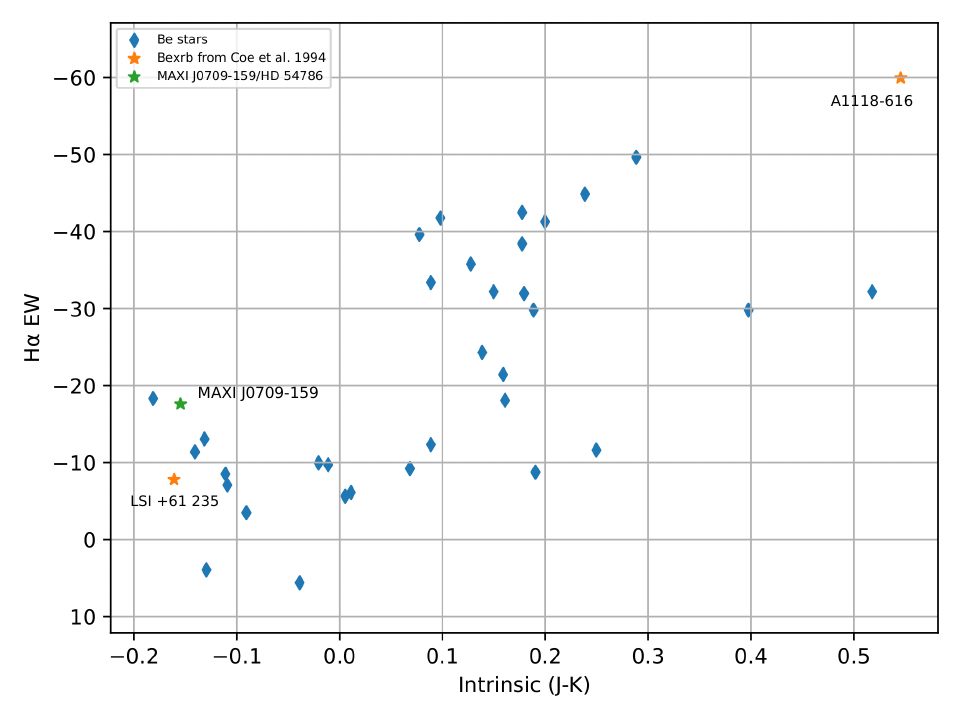}
\bigskip
\begin{minipage}{12cm}
\caption{Plot of H$\alpha$ versus intrinsic J-K for MAXI J0709-159 observed in February 1, 2022 (Green star symbol). The location of other two HMXB are marked in orange marker. The blue diamond symbols are for the field Be stars from Dachs and ashoke. The diagram indicates that the location of HMXB (BeXRB in this case) can differ vastly based on the state of the CSM of the particular system. In case of MAXI J0709-159, it is obtaining comparatively smaller disc size.}
    \label{figure2} 
    \end{minipage}
\end{figure}

We then performed an analysis involving the H$\alpha$ and extinction corrected (J-K) plot to gain a photometric perspective on the circumstellar disk around the primary object in HMXB systems. This kind of analysis has been extensively conducted by \cite[]{1993Coe, 1994Coe} to explore the properties of the BeXRB objects LSI +61$^{\circ}$ 235 and A1118-616. Here, we utilize the same plot to understand the location of the optical counter part (HD 54786) of the X-ray source MAXI J0709-159 among the candidate sources. In Fig.\,\ref{figure2}, we present the H$\alpha$ vs intrinsic (J-K) plot for a sample of field Be stars \citep[]{1982Dachs, 1984Ashok} along with two BeXRB objects (mentioned above). The H$\alpha$ equivalent width (EW) for HD 54786 was measured to be -16.9 \AA, as reported in \cite{2022Bhattacharyya}. The observations were conducted using the Himalayan Chandra telescope (HCT) located at the Indian Astronomical Observatory (IAO), Ladakh, India on February 1, 2022, precisely 6 days after the reported flare \citep{2022ATelserino}. IAO is operated by the Indian Institute of Astrophysics (IIA), Bangalore. On the other hand, the J and K magnitudes were obtained from the 2MASS catalog \citep{2003Cutri}. These magnitudes were corrected using the values provided by \cite{2019Green} to account for extinction.

In Fig.\,\ref{figure2}, the orange star symbol represents the location of the two BeXRB objects from \cite[]{1993Coe, 1994Coe}. The green star symbol indicates the location of HD 54786 (MAXI J0709-159), while the blue diamond symbol depicts the location of field Be stars taken from \cite{1982Dachs} and \cite{1984Ashok}. The figure clearly illustrates that the positions of HMXBs, specifically the two BeXRBs in this case, can vary significantly based on the state of the circumstellar medium (CSM) of each system.

\cite{1993Coe} proposed that the CSM of LSI +61$^{\circ}$ 235, being of low density and/or temperature, results in negligible energy in the near-IR region, leading to the conclusion that the system exhibits low-level activity during that time and has little material around the Be star. Similarly to HD 54786 (MAXI J0709-159), which is located close to LSI +61$^{\circ}$ 235 in the H$\alpha$ vs. (J-K) plot, our object may also have a scarcity of material around the Be star and experience a low level of activity, as the H$\alpha$ observation took place 6 days after the flare. A weak infrared (IR) excess may be attributed to free-free emission in the disc or shell structure surrounding a Be star, which is a common observation in both Be stars and several X-ray systems associated with Be stars \citep{1988Coe}.

\section{Conclusions}

In our previous study \citep{2022Bhattacharyya}, we established HD 54786 as an evolved star departing from the main sequence by comparing with non-X-ray binary systems. In this current study, the CMD analysis based on HMXB, coupled with t-tests, increases the probability that HD 54786 maybe a member of both the "Bexrb" and "Sgxrb" populations of HMXBs, signifying its dual optical characteristics as an X-ray binary star and its distinct evolutionary phase from BeXRB towards SgXB stage.

We also utilized the H$\alpha$ vs J-K plot to gain insights into the circumstellar disk around the primary object in HMXB systems, the current analysis indicates that the system MAXI J0709-159 is currently exhibiting low-level activity, with less amount of material surrounding the Be star HD 54786. The current state of our object shows similarities to the study on LSI +61$^{\circ}$ 235 \cite{1994Coe}, although that system benefited from extensive monitoring data, which proved extremely valuable in understanding the behavior of this intricate system. Continued monitoring and data collection are crucial in unraveling the complexities of MAXI J0709-159 and its evolution within the context of HMXBs.

\begin{furtherinformation}

\begin{orcids}
\orcid{0000-0002-1920-6055}{Suman}{Bhattacharyya}
\orcid{0000-0002-7254-191X}{Blesson}{Mathew}
\orcid{0000-0001-8873-1171}{Gourav}{Banerjee}
\orcid{0000-0002-4024-956X}{S.}{Muneer}
\orcid{0009-0007-1545-854X}{Santosh}{Joshi}
\end{orcids}

\begin{authorcontributions}
This work is part of a collective effort with contributions from all the co-authors.
\end{authorcontributions}

\begin{conflictsofinterest}
The authors declare no conflict of interest.
\end{conflictsofinterest}

\end{furtherinformation}

\section{Acknowledgement}
We convey our heartiest gratitude to the staff of the Indian Astronomical Observatory (IAO), Ladakh for taking the observations, respectively. We would like to thank the Center for research, CHRIST (Deemed to be University), Bangalore, India for providing the necessary support. R.A is grateful to the Centre for Research, CHRIST (Deemed to be University), Bangalore for the research grant extended to carry out the present project (MRP DSC-1932). BM acknowledges the support of the Science \& Engineering Research Board (SERB), a statutory body of the Department of Science \& Technology (DST), Government of India, for funding our research under grant number CRG/2019/005380. Finally, SB would like to acknowledge the local support provided by the BINA project during the time of BINA-2023 workshop.


\bibliographystyle{bullsrsl-en}

\bibliography{S08-CT01_BhattacharyyaS}

\begin{thebibliography}{29}
\providecommand{\natexlab}[1]{#1}
\providecommand{\url}[1]{#1}
\providecommand{\urlprefix}{URL }

\bibitem[{{Ashok} et~al.(1984){Ashok}, {Bhatt}, {Kulkarni} and
  {Joshi}}]{1984Ashok}
{Ashok}, N.~M., {Bhatt}, H.~C., {Kulkarni}, P.~V. and {Joshi}, S.~C. (1984)
  {Infrared photometric studies of Be stars.}
\newblock MNRAS, 211, 471--484.
\newblock \url{https://doi.org/10.1093/mnras/211.2.471}.

\bibitem[{{Asplund} et~al.(2009){Asplund}, {Grevesse}, {Sauval} and
  {Scott}}]{2009Asplund}
{Asplund}, M., {Grevesse}, N., {Sauval}, A.~J. and {Scott}, P. (2009) {The
  Chemical Composition of the Sun}.
\newblock ARA\&A, 47(1), 481--522.
\newblock \url{https://doi.org/10.1146/annurev.astro.46.060407.145222}.

\bibitem[{{Bailer-Jones} et~al.(2021){Bailer-Jones}, {Rybizki}, {Fouesneau},
  {Demleitner} and {Andrae}}]{2021Bailerjonesdr3}
{Bailer-Jones}, C.~A.~L., {Rybizki}, J., {Fouesneau}, M., {Demleitner}, M. and
  {Andrae}, R. (2021) {Estimating Distances from Parallaxes. V. Geometric and
  Photogeometric Distances to 1.47 Billion Stars in Gaia Early Data Release 3}.
\newblock AJ, 161(3), 147.
\newblock \url{https://doi.org/10.3847/1538-3881/abd806}.

\bibitem[{{Belczynski} and {Ziolkowski}(2009)}]{2009Belczynski}
{Belczynski}, K. and {Ziolkowski}, J. (2009) {On the Apparent Lack of Be X-Ray
  Binaries with Black Holes}.
\newblock ApJ, 707(2), 870--877.
\newblock \url{https://doi.org/10.1088/0004-637X/707/2/870}.

\bibitem[{{Bhattacharyya} et~al.(2021){Bhattacharyya}, {Mathew}, {Banerjee},
  {Anusha}, {Paul} and {Kartha}}]{2021Bhattacharyya}
{Bhattacharyya}, S., {Mathew}, B., {Banerjee}, G., {Anusha}, R., {Paul}, K.~T.
  and {Kartha}, S.~S. (2021) {Identification of emission-line stars in
  transition phase from pre-main sequence to main sequence}.
\newblock MNRAS, 507(3), 3660--3671.
\newblock \url{https://doi.org/10.1093/mnras/stab2385}.

\bibitem[{{Bhattacharyya} et~al.(2022){Bhattacharyya}, {Mathew}, {Ezhikode},
  {Muneer}, {Selvakumar}, {Maheswer}, {Arun}, {Anilkumar}, {Banerjee},
  {Pramod}, {Kartha}, {Paul} and {Velu}}]{2022Bhattacharyya}
{Bhattacharyya}, S., {Mathew}, B., {Ezhikode}, S.~H., {Muneer}, S.,
  {Selvakumar}, G., {Maheswer}, G., {Arun}, R., {Anilkumar}, H., {Banerjee},
  G., {Pramod}, K.~S., {Kartha}, S.~S., {Paul}, K.~T. and {Velu}, C. (2022)
  {Decoding the X-Ray Flare from MAXI J0709-159 Using Optical Spectroscopy and
  Multiepoch Photometry}.
\newblock ApJ, 933(2), L34.
\newblock \url{https://doi.org/10.3847/2041-8213/ac7b8a}.

\bibitem[{{Choi} et~al.(2016){Choi}, {Dotter}, {Conroy}, {Cantiello}, {Paxton}
  and {Johnson}}]{2016choi}
{Choi}, J., {Dotter}, A., {Conroy}, C., {Cantiello}, M., {Paxton}, B. and
  {Johnson}, B.~D. (2016) {Mesa Isochrones and Stellar Tracks (MIST). I.
  Solar-scaled Models}.
\newblock ApJ, 823(2), 102.
\newblock \url{https://doi.org/10.3847/0004-637X/823/2/102}.

\bibitem[{{Coe} et~al.(1993){Coe}, {Everall}, {Norton}, {Roche}, {Unger},
  {Fabregat}, {Reglero} and {Grunsfeld}}]{1993Coe}
{Coe}, M.~J., {Everall}, C., {Norton}, A.~J., {Roche}, P., {Unger}, S.~J.,
  {Fabregat}, J., {Reglero}, V. and {Grunsfeld}, J.~M. (1993) {Infrared and
  optical observations of the newly identified Be/X-ray binary LSI +61 235.}
\newblock MNRAS, 261, 599--604.
\newblock \url{https://doi.org/10.1093/mnras/261.3.599}.

\bibitem[{{Coe} et~al.(1988){Coe}, {Longmore}, {Payne} and {Hanson}}]{1988Coe}
{Coe}, M.~J., {Longmore}, A., {Payne}, B.~J. and {Hanson}, C.~G. (1988) {The
  optical/IR counterpart to the newly-discovered X-ray source EXO 2030+375.}
\newblock MNRAS, 232, 865--871.
\newblock \url{https://doi.org/10.1093/mnras/232.4.865}.

\bibitem[{{Coe} et~al.(1994){Coe}, {Roche}, {Everall}, {Fishman}, {Hagedon},
  {Finger}, {Wilson}, {Buckley}, {Shrader}, {Fabregat}, {Polcaro},
  {Giovannelli} and {Villada}}]{1994Coe}
{Coe}, M.~J., {Roche}, P., {Everall}, C., {Fishman}, G.~J., {Hagedon}, K.~S.,
  {Finger}, M., {Wilson}, R.~B., {Buckley}, D.~A.~H., {Shrader}, C.,
  {Fabregat}, J., {Polcaro}, V.~F., {Giovannelli}, F. and {Villada}, M. (1994)
  {Multiwaveband study of a major X-ray outburst from the Be/X-ray transient
  system A 1118-616.}
\newblock A\&A, 289, 784--794.

\bibitem[{{Cutri} et~al.(2003){Cutri}, {Skrutskie}, {van Dyk}, {Beichman},
  {Carpenter}, {Chester}, {Cambresy}, {Evans}, {Fowler}, {Gizis}, {Howard},
  {Huchra}, {Jarrett}, {Kopan}, {Kirkpatrick}, {Light}, {Marsh}, {McCallon},
  {Schneider}, {Stiening}, {Sykes}, {Weinberg}, {Wheaton}, {Wheelock} and
  {Zacarias}}]{2003Cutri}
{Cutri}, R.~M., {Skrutskie}, M.~F., {van Dyk}, S., {Beichman}, C.~A.,
  {Carpenter}, J.~M., {Chester}, T., {Cambresy}, L., {Evans}, T., {Fowler}, J.,
  {Gizis}, J., {Howard}, E., {Huchra}, J., {Jarrett}, T., {Kopan}, E.~L.,
  {Kirkpatrick}, J.~D., {Light}, R.~M., {Marsh}, K.~A., {McCallon}, H.,
  {Schneider}, S., {Stiening}, R., {Sykes}, M., {Weinberg}, M., {Wheaton},
  W.~A., {Wheelock}, S. and {Zacarias}, N. (2003) {VizieR Online Data Catalog:
  2MASS All-Sky Catalog of Point Sources (Cutri+ 2003)}.
\newblock VizieR Online Data Catalog, II/246.

\bibitem[{{Dachs} and {Wamsteker}(1982)}]{1982Dachs}
{Dachs}, J. and {Wamsteker}, W. (1982) {Infrared photometry of southern Be
  stars.}
\newblock A\&A, 107, 240--246.

\bibitem[{{Dotter}(2016)}]{2016dotter}
{Dotter}, A. (2016) {MESA Isochrones and Stellar Tracks (MIST) 0: Methods for
  the Construction of Stellar Isochrones}.
\newblock ApJS, 222(1), 8.
\newblock \url{https://doi.org/10.3847/0067-0049/222/1/8}.

\bibitem[{{Drave}(2013)}]{drave2013}
{Drave}, S. (2013) {Supergiant fast X-ray transients}.
\newblock A\&G, 54(6), 6.27--6.30.
\newblock \url{https://doi.org/10.1093/astrogeo/att204}.

\bibitem[{{Gaia Collaboration} et~al.(2021){Gaia Collaboration}, {Brown},
  {Vallenari}, {Prusti}, {de Bruijne}, {Babusiaux}, {Biermann}, {Creevey},
  {Evans}, {Eyer}, {Hutton}, {Jansen}, {Jordi}, {Klioner}, {Lammers},
  {Lindegren}, {Luri}, {Mignard}, {Panem}, {Pourbaix}, {Randich}, {Sartoretti},
  {Soubiran}, {Walton}, {Arenou}, {Bailer-Jones}, {Bastian}, {Cropper},
  {Drimmel}, {Katz}, {Lattanzi}, {van Leeuwen}, {Bakker}, {Cacciari},
  {Casta{\~n}eda}, {De Angeli}, {Ducourant}, {Fabricius}, {Fouesneau},
  {Fr{\'e}mat}, {Guerra}, {Guerrier}, {Guiraud}, {Jean-Antoine Piccolo},
  {Masana}, {Messineo}, {Mowlavi}, {Nicolas}, {Nienartowicz}, {Pailler},
  {Panuzzo}, {Riclet}, {Roux}, {Seabroke}, {Sordo}, {Tanga}, {Th{\'e}venin},
  {Gracia-Abril}, {Portell}, {Teyssier}, {Altmann}, {Andrae}, {Bellas-Velidis},
  {Benson}, {Berthier}, {Blomme}, {Brugaletta}, {Burgess}, {Busso}, {Carry},
  {Cellino}, {Cheek}, {Clementini}, {Damerdji}, {Davidson}, {Delchambre},
  {Dell'Oro}, {Fern{\'a}ndez-Hern{\'a}ndez}, {Galluccio}, {Garc{\'\i}a-Lario},
  {Garcia-Reinaldos}, {Gonz{\'a}lez-N{\'u}{\~n}ez}, {Gosset}, {Haigron},
  {Halbwachs}, {Hambly}, {Harrison}, {Hatzidimitriou}, {Heiter},
  {Hern{\'a}ndez}, {Hestroffer}, {Hodgkin}, {Holl}, {Jan{\ss}en}, {Jevardat de
  Fombelle}, {Jordan}, {Krone-Martins}, {Lanzafame}, {L{\"o}ffler}, {Lorca},
  {Manteiga}, {Marchal}, {Marrese}, {Moitinho}, {Mora}, {Muinonen}, {Osborne},
  {Pancino}, {Pauwels}, {Petit}, {Recio-Blanco}, {Richards}, {Riello},
  {Rimoldini}, {Robin}, {Roegiers}, {Rybizki}, {Sarro}, {Siopis}, {Smith},
  {Sozzetti}, {Ulla}, {Utrilla}, {van Leeuwen}, {van Reeven}, {Abbas}, {Abreu
  Aramburu}, {Accart}, {Aerts}, {Aguado}, {Ajaj}, {Altavilla}, {{\'A}lvarez},
  {{\'A}lvarez Cid-Fuentes}, {Alves}, {Anderson}, {Anglada Varela}, {Antoja},
  {Audard}, {Baines}, {Baker}, {Balaguer-N{\'u}{\~n}ez}, {Balbinot}, {Balog},
  {Barache}, {Barbato}, {Barros}, {Barstow}, {Bartolom{\'e}}, {Bassilana},
  {Bauchet}, {Baudesson-Stella}, {Becciani}, {Bellazzini}, {Bernet}, {Bertone},
  {Bianchi}, {Blanco-Cuaresma}, {Boch}, {Bombrun}, {Bossini}, {Bouquillon},
  {Bragaglia}, {Bramante}, {Breedt}, {Bressan}, {Brouillet}, {Bucciarelli},
  {Burlacu}, {Busonero}, {Butkevich}, {Buzzi}, {Caffau}, {Cancelliere},
  {C{\'a}novas}, {Cantat-Gaudin}, {Carballo}, {Carlucci}, {Carnerero},
  {Carrasco}, {Casamiquela}, {Castellani}, {Castro-Ginard}, {Castro Sampol},
  {Chaoul}, {Charlot}, {Chemin}, {Chiavassa}, {Cioni}, {Comoretto}, {Cooper},
  {Cornez}, {Cowell}, {Crifo}, {Crosta}, {Crowley}, {Dafonte}, {Dapergolas},
  {David}, {David}, {de Laverny}, {De Luise}, {De March}, {De Ridder}, {de
  Souza}, {de Teodoro}, {de Torres}, {del Peloso}, {del Pozo}, {Delbo},
  {Delgado}, {Delgado}, {Delisle}, {Di Matteo}, {Diakite}, {Diener},
  {Distefano}, {Dolding}, {Eappachen}, {Edvardsson}, {Enke}, {Esquej}, {Fabre},
  {Fabrizio}, {Faigler}, {Fedorets}, {Fernique}, {Fienga}, {Figueras},
  {Fouron}, {Fragkoudi}, {Fraile}, {Franke}, {Gai}, {Garabato},
  {Garcia-Gutierrez}, {Garc{\'\i}a-Torres}, {Garofalo}, {Gavras}, {Gerlach},
  {Geyer}, {Giacobbe}, {Gilmore}, {Girona}, {Giuffrida}, {Gomel}, {Gomez},
  {Gonzalez-Santamaria}, {Gonz{\'a}lez-Vidal}, {Granvik},
  {Guti{\'e}rrez-S{\'a}nchez}, {Guy}, {Hauser}, {Haywood}, {Helmi}, {Hidalgo},
  {Hilger}, {H{\l}adczuk}, {Hobbs}, {Holland}, {Huckle}, {Jasniewicz},
  {Jonker}, {Juaristi Campillo}, {Julbe}, {Karbevska}, {Kervella}, {Khanna},
  {Kochoska}, {Kontizas}, {Kordopatis}, {Korn}, {Kostrzewa-Rutkowska},
  {Kruszy{\'n}ska}, {Lambert}, {Lanza}, {Lasne}, {Le Campion}, {Le Fustec},
  {Lebreton}, {Lebzelter}, {Leccia}, {Leclerc}, {Lecoeur-Taibi}, {Liao},
  {Licata}, {Lindstr{\o}m}, {Lister}, {Livanou}, {Lobel}, {Madrero Pardo},
  {Managau}, {Mann}, {Marchant}, {Marconi}, {Marcos Santos}, {Marinoni},
  {Marocco}, {Marshall}, {Martin Polo}, {Mart{\'\i}n-Fleitas}, {Masip},
  {Massari}, {Mastrobuono-Battisti}, {Mazeh}, {McMillan}, {Messina},
  {Michalik}, {Millar}, {Mints}, {Molina}, {Molinaro}, {Moln{\'a}r},
  {Montegriffo}, {Mor}, {Morbidelli}, {Morel}, {Morris}, {Mulone}, {Munoz},
  {Muraveva}, {Murphy}, {Musella}, {Noval}, {Ord{\'e}novic}, {Orr{\`u}},
  {Osinde}, {Pagani}, {Pagano}, {Palaversa}, {Palicio}, {Panahi}, {Pawlak},
  {Pe{\~n}alosa Esteller}, {Penttil{\"a}}, {Piersimoni}, {Pineau}, {Plachy},
  {Plum}, {Poggio}, {Poretti}, {Poujoulet}, {Pr{\v{s}}a}, {Pulone}, {Racero},
  {Ragaini}, {Rainer}, {Raiteri}, {Rambaux}, {Ramos}, {Ramos-Lerate}, {Re
  Fiorentin}, {Regibo}, {Reyl{\'e}}, {Ripepi}, {Riva}, {Rixon}, {Robichon},
  {Robin}, {Roelens}, {Rohrbasser}, {Romero-G{\'o}mez}, {Rowell}, {Royer},
  {Rybicki}, {Sadowski}, {Sagrist{\`a} Sell{\'e}s}, {Sahlmann}, {Salgado},
  {Salguero}, {Samaras}, {Sanchez Gimenez}, {Sanna}, {Santove{\~n}a},
  {Sarasso}, {Schultheis}, {Sciacca}, {Segol}, {Segovia}, {S{\'e}gransan},
  {Semeux}, {Shahaf}, {Siddiqui}, {Siebert}, {Siltala}, {Slezak}, {Smart},
  {Solano}, {Solitro}, {Souami}, {Souchay}, {Spagna}, {Spoto}, {Steele},
  {Steidelm{\"u}ller}, {Stephenson}, {S{\"u}veges}, {Szabados}, {Szegedi-Elek},
  {Taris}, {Tauran}, {Taylor}, {Teixeira}, {Thuillot}, {Tonello}, {Torra},
  {Torra}, {Turon}, {Unger}, {Vaillant}, {van Dillen}, {Vanel}, {Vecchiato},
  {Viala}, {Vicente}, {Voutsinas}, {Weiler}, {Wevers}, {Wyrzykowski}, {Yoldas},
  {Yvard}, {Zhao}, {Zorec}, {Zucker}, {Zurbach} and
  {Zwitter}}]{Gaia2021A&A...649A...1G}
{Gaia Collaboration}, {Brown}, A.~G.~A., {Vallenari}, A., {Prusti}, T., {de
  Bruijne}, J.~H.~J., {Babusiaux}, C., {Biermann}, M., {Creevey}, O.~L.,
  {Evans}, D.~W., {Eyer}, L., {Hutton}, A., {Jansen}, F., {Jordi}, C.,
  {Klioner}, S.~A., {Lammers}, U., {Lindegren}, L., {Luri}, X., {Mignard}, F.,
  {Panem}, C., {Pourbaix}, D., {Randich}, S., {Sartoretti}, P., {Soubiran}, C.,
  {Walton}, N.~A., {Arenou}, F., {Bailer-Jones}, C.~A.~L., {Bastian}, U.,
  {Cropper}, M., {Drimmel}, R., {Katz}, D., {Lattanzi}, M.~G., {van Leeuwen},
  F., {Bakker}, J., {Cacciari}, C., {Casta{\~n}eda}, J., {De Angeli}, F.,
  {Ducourant}, C., {Fabricius}, C., {Fouesneau}, M., {Fr{\'e}mat}, Y.,
  {Guerra}, R., {Guerrier}, A., {Guiraud}, J., {Jean-Antoine Piccolo}, A.,
  {Masana}, E., {Messineo}, R., {Mowlavi}, N., {Nicolas}, C., {Nienartowicz},
  K., {Pailler}, F., {Panuzzo}, P., {Riclet}, F., {Roux}, W., {Seabroke},
  G.~M., {Sordo}, R., {Tanga}, P., {Th{\'e}venin}, F., {Gracia-Abril}, G.,
  {Portell}, J., {Teyssier}, D., {Altmann}, M., {Andrae}, R., {Bellas-Velidis},
  I., {Benson}, K., {Berthier}, J., {Blomme}, R., {Brugaletta}, E., {Burgess},
  P.~W., {Busso}, G., {Carry}, B., {Cellino}, A., {Cheek}, N., {Clementini},
  G., {Damerdji}, Y., {Davidson}, M., {Delchambre}, L., {Dell'Oro}, A.,
  {Fern{\'a}ndez-Hern{\'a}ndez}, J., {Galluccio}, L., {Garc{\'\i}a-Lario}, P.,
  {Garcia-Reinaldos}, M., {Gonz{\'a}lez-N{\'u}{\~n}ez}, J., {Gosset}, E.,
  {Haigron}, R., {Halbwachs}, J.~L., {Hambly}, N.~C., {Harrison}, D.~L.,
  {Hatzidimitriou}, D., {Heiter}, U., {Hern{\'a}ndez}, J., {Hestroffer}, D.,
  {Hodgkin}, S.~T., {Holl}, B., {Jan{\ss}en}, K., {Jevardat de Fombelle}, G.,
  {Jordan}, S., {Krone-Martins}, A., {Lanzafame}, A.~C., {L{\"o}ffler}, W.,
  {Lorca}, A., {Manteiga}, M., {Marchal}, O., {Marrese}, P.~M., {Moitinho}, A.,
  {Mora}, A., {Muinonen}, K., {Osborne}, P., {Pancino}, E., {Pauwels}, T.,
  {Petit}, J.~M., {Recio-Blanco}, A., {Richards}, P.~J., {Riello}, M.,
  {Rimoldini}, L., {Robin}, A.~C., {Roegiers}, T., {Rybizki}, J., {Sarro},
  L.~M., {Siopis}, C., {Smith}, M., {Sozzetti}, A., {Ulla}, A., {Utrilla}, E.,
  {van Leeuwen}, M., {van Reeven}, W., {Abbas}, U., {Abreu Aramburu}, A.,
  {Accart}, S., {Aerts}, C., {Aguado}, J.~J., {Ajaj}, M., {Altavilla}, G.,
  {{\'A}lvarez}, M.~A., {{\'A}lvarez Cid-Fuentes}, J., {Alves}, J., {Anderson},
  R.~I., {Anglada Varela}, E., {Antoja}, T., {Audard}, M., {Baines}, D.,
  {Baker}, S.~G., {Balaguer-N{\'u}{\~n}ez}, L., {Balbinot}, E., {Balog}, Z.,
  {Barache}, C., {Barbato}, D., {Barros}, M., {Barstow}, M.~A.,
  {Bartolom{\'e}}, S., {Bassilana}, J.~L., {Bauchet}, N., {Baudesson-Stella},
  A., {Becciani}, U., {Bellazzini}, M., {Bernet}, M., {Bertone}, S., {Bianchi},
  L., {Blanco-Cuaresma}, S., {Boch}, T., {Bombrun}, A., {Bossini}, D.,
  {Bouquillon}, S., {Bragaglia}, A., {Bramante}, L., {Breedt}, E., {Bressan},
  A., {Brouillet}, N., {Bucciarelli}, B., {Burlacu}, A., {Busonero}, D.,
  {Butkevich}, A.~G., {Buzzi}, R., {Caffau}, E., {Cancelliere}, R.,
  {C{\'a}novas}, H., {Cantat-Gaudin}, T., {Carballo}, R., {Carlucci}, T.,
  {Carnerero}, M.~I., {Carrasco}, J.~M., {Casamiquela}, L., {Castellani}, M.,
  {Castro-Ginard}, A., {Castro Sampol}, P., {Chaoul}, L., {Charlot}, P.,
  {Chemin}, L., {Chiavassa}, A., {Cioni}, M. R.~L., {Comoretto}, G., {Cooper},
  W.~J., {Cornez}, T., {Cowell}, S., {Crifo}, F., {Crosta}, M., {Crowley}, C.,
  {Dafonte}, C., {Dapergolas}, A., {David}, M., {David}, P., {de Laverny}, P.,
  {De Luise}, F., {De March}, R., {De Ridder}, J., {de Souza}, R., {de
  Teodoro}, P., {de Torres}, A., {del Peloso}, E.~F., {del Pozo}, E., {Delbo},
  M., {Delgado}, A., {Delgado}, H.~E., {Delisle}, J.~B., {Di Matteo}, P.,
  {Diakite}, S., {Diener}, C., {Distefano}, E., {Dolding}, C., {Eappachen}, D.,
  {Edvardsson}, B., {Enke}, H., {Esquej}, P., {Fabre}, C., {Fabrizio}, M.,
  {Faigler}, S., {Fedorets}, G., {Fernique}, P., {Fienga}, A., {Figueras}, F.,
  {Fouron}, C., {Fragkoudi}, F., {Fraile}, E., {Franke}, F., {Gai}, M.,
  {Garabato}, D., {Garcia-Gutierrez}, A., {Garc{\'\i}a-Torres}, M., {Garofalo},
  A., {Gavras}, P., {Gerlach}, E., {Geyer}, R., {Giacobbe}, P., {Gilmore}, G.,
  {Girona}, S., {Giuffrida}, G., {Gomel}, R., {Gomez}, A.,
  {Gonzalez-Santamaria}, I., {Gonz{\'a}lez-Vidal}, J.~J., {Granvik}, M.,
  {Guti{\'e}rrez-S{\'a}nchez}, R., {Guy}, L.~P., {Hauser}, M., {Haywood}, M.,
  {Helmi}, A., {Hidalgo}, S.~L., {Hilger}, T., {H{\l}adczuk}, N., {Hobbs}, D.,
  {Holland}, G., {Huckle}, H.~E., {Jasniewicz}, G., {Jonker}, P.~G., {Juaristi
  Campillo}, J., {Julbe}, F., {Karbevska}, L., {Kervella}, P., {Khanna}, S.,
  {Kochoska}, A., {Kontizas}, M., {Kordopatis}, G., {Korn}, A.~J.,
  {Kostrzewa-Rutkowska}, Z., {Kruszy{\'n}ska}, K., {Lambert}, S., {Lanza},
  A.~F., {Lasne}, Y., {Le Campion}, J.~F., {Le Fustec}, Y., {Lebreton}, Y.,
  {Lebzelter}, T., {Leccia}, S., {Leclerc}, N., {Lecoeur-Taibi}, I., {Liao},
  S., {Licata}, E., {Lindstr{\o}m}, E.~P., {Lister}, T.~A., {Livanou}, E.,
  {Lobel}, A., {Madrero Pardo}, P., {Managau}, S., {Mann}, R.~G., {Marchant},
  J.~M., {Marconi}, M., {Marcos Santos}, M.~M.~S., {Marinoni}, S., {Marocco},
  F., {Marshall}, D.~J., {Martin Polo}, L., {Mart{\'\i}n-Fleitas}, J.~M.,
  {Masip}, A., {Massari}, D., {Mastrobuono-Battisti}, A., {Mazeh}, T.,
  {McMillan}, P.~J., {Messina}, S., {Michalik}, D., {Millar}, N.~R., {Mints},
  A., {Molina}, D., {Molinaro}, R., {Moln{\'a}r}, L., {Montegriffo}, P., {Mor},
  R., {Morbidelli}, R., {Morel}, T., {Morris}, D., {Mulone}, A.~F., {Munoz},
  D., {Muraveva}, T., {Murphy}, C.~P., {Musella}, I., {Noval}, L.,
  {Ord{\'e}novic}, C., {Orr{\`u}}, G., {Osinde}, J., {Pagani}, C., {Pagano},
  I., {Palaversa}, L., {Palicio}, P.~A., {Panahi}, A., {Pawlak}, M.,
  {Pe{\~n}alosa Esteller}, X., {Penttil{\"a}}, A., {Piersimoni}, A.~M.,
  {Pineau}, F.~X., {Plachy}, E., {Plum}, G., {Poggio}, E., {Poretti}, E.,
  {Poujoulet}, E., {Pr{\v{s}}a}, A., {Pulone}, L., {Racero}, E., {Ragaini}, S.,
  {Rainer}, M., {Raiteri}, C.~M., {Rambaux}, N., {Ramos}, P., {Ramos-Lerate},
  M., {Re Fiorentin}, P., {Regibo}, S., {Reyl{\'e}}, C., {Ripepi}, V., {Riva},
  A., {Rixon}, G., {Robichon}, N., {Robin}, C., {Roelens}, M., {Rohrbasser},
  L., {Romero-G{\'o}mez}, M., {Rowell}, N., {Royer}, F., {Rybicki}, K.~A.,
  {Sadowski}, G., {Sagrist{\`a} Sell{\'e}s}, A., {Sahlmann}, J., {Salgado}, J.,
  {Salguero}, E., {Samaras}, N., {Sanchez Gimenez}, V., {Sanna}, N.,
  {Santove{\~n}a}, R., {Sarasso}, M., {Schultheis}, M., {Sciacca}, E., {Segol},
  M., {Segovia}, J.~C., {S{\'e}gransan}, D., {Semeux}, D., {Shahaf}, S.,
  {Siddiqui}, H.~I., {Siebert}, A., {Siltala}, L., {Slezak}, E., {Smart},
  R.~L., {Solano}, E., {Solitro}, F., {Souami}, D., {Souchay}, J., {Spagna},
  A., {Spoto}, F., {Steele}, I.~A., {Steidelm{\"u}ller}, H., {Stephenson},
  C.~A., {S{\"u}veges}, M., {Szabados}, L., {Szegedi-Elek}, E., {Taris}, F.,
  {Tauran}, G., {Taylor}, M.~B., {Teixeira}, R., {Thuillot}, W., {Tonello}, N.,
  {Torra}, F., {Torra}, J., {Turon}, C., {Unger}, N., {Vaillant}, M., {van
  Dillen}, E., {Vanel}, O., {Vecchiato}, A., {Viala}, Y., {Vicente}, D.,
  {Voutsinas}, S., {Weiler}, M., {Wevers}, T., {Wyrzykowski}, {\L}., {Yoldas},
  A., {Yvard}, P., {Zhao}, H., {Zorec}, J., {Zucker}, S., {Zurbach}, C. and
  {Zwitter}, T. (2021) {Gaia Early Data Release 3. Summary of the contents and
  survey properties}.
\newblock A\&A, 649, A1.
\newblock \url{https://doi.org/10.1051/0004-6361/202039657}.

\bibitem[{{Georgy} et~al.(2021){Georgy}, {Saio} and {Meynet}}]{2021Georgy}
{Georgy}, C., {Saio}, H. and {Meynet}, G. (2021) {Blue supergiants as tests for
  stellar physics}.
\newblock A\&A, 650, A128.
\newblock \url{https://doi.org/10.1051/0004-6361/202040105}.

\bibitem[{{Green} et~al.(2019){Green}, {Schlafly}, {Zucker}, {Speagle} and
  {Finkbeiner}}]{2019Green}
{Green}, G.~M., {Schlafly}, E., {Zucker}, C., {Speagle}, J.~S. and
  {Finkbeiner}, D. (2019) {A 3D Dust Map Based on Gaia, Pan-STARRS 1, and
  2MASS}.
\newblock ApJ, 887(1), 93.
\newblock \url{https://doi.org/10.3847/1538-4357/ab5362}.

\bibitem[{{Hohle} et~al.(2010){Hohle}, {Neuh{\"a}user} and
  {Schutz}}]{Hohle2010AN....331..349H}
{Hohle}, M.~M., {Neuh{\"a}user}, R. and {Schutz}, B.~F. (2010) {Masses and
  luminosities of O- and B-type stars and red supergiants}.
\newblock AN, 331(4), 349.
\newblock \url{https://doi.org/10.1002/asna.200911355}.

\bibitem[{{Huang} et~al.(2010){Huang}, {Gies} and {McSwain}}]{2010Huang}
{Huang}, W., {Gies}, D.~R. and {McSwain}, M.~V. (2010) {A Stellar Rotation
  Census of B Stars: From ZAMS to TAMS}.
\newblock ApJ, 722(1), 605--619.
\newblock \url{https://doi.org/10.1088/0004-637X/722/1/605}.

\bibitem[{{Kretschmar} et~al.(2019){Kretschmar}, {F{\"u}rst}, {Sidoli},
  {Bozzo}, {Alfonso-Garz{\'o}n}, {Bodaghee}, {Chaty}, {Chernyakova},
  {Ferrigno}, {Manousakis}, {Negueruela}, {Postnov}, {Paizis}, {Reig},
  {Rodes-Roca}, {Tsygankov}, {Bird}, {Bissinger n{\'e} K{\"u}hnel}, {Blay},
  {Caballero}, {Coe}, {Domingo}, {Doroshenko}, {Ducci}, {Falanga}, {Grebenev},
  {Grinberg}, {Hemphill}, {Kreykenbohm}, {Kreykenbohm n{\'e} Fritz}, {Li},
  {Lutovinov}, {Mart{\'\i}nez-N{\'u}{\~n}ez}, {Mas-Hesse}, {Masetti},
  {McBride}, {Neronov}, {Pottschmidt}, {Rodriguez}, {Romano}, {Rothschild},
  {Santangelo}, {Sguera}, {Staubert}, {Tomsick}, {Torrej{\'o}n}, {Torres},
  {Walter}, {Wilms}, {Wilson-Hodge} and {Zhang}}]{2019Kretschmar}
{Kretschmar}, P., {F{\"u}rst}, F., {Sidoli}, L., {Bozzo}, E.,
  {Alfonso-Garz{\'o}n}, J., {Bodaghee}, A., {Chaty}, S., {Chernyakova}, M.,
  {Ferrigno}, C., {Manousakis}, A., {Negueruela}, I., {Postnov}, K., {Paizis},
  A., {Reig}, P., {Rodes-Roca}, J.~J., {Tsygankov}, S., {Bird}, A.~J.,
  {Bissinger n{\'e} K{\"u}hnel}, M., {Blay}, P., {Caballero}, I., {Coe}, M.~J.,
  {Domingo}, A., {Doroshenko}, V., {Ducci}, L., {Falanga}, M., {Grebenev},
  S.~A., {Grinberg}, V., {Hemphill}, P., {Kreykenbohm}, I., {Kreykenbohm n{\'e}
  Fritz}, S., {Li}, J., {Lutovinov}, A.~A., {Mart{\'\i}nez-N{\'u}{\~n}ez}, S.,
  {Mas-Hesse}, J.~M., {Masetti}, N., {McBride}, V.~A., {Neronov}, A.,
  {Pottschmidt}, K., {Rodriguez}, J., {Romano}, P., {Rothschild}, R.~E.,
  {Santangelo}, A., {Sguera}, V., {Staubert}, R., {Tomsick}, J.~A.,
  {Torrej{\'o}n}, J.~M., {Torres}, D.~F., {Walter}, R., {Wilms}, J.,
  {Wilson-Hodge}, C.~A. and {Zhang}, S. (2019) {Advances in Understanding
  High-Mass X-ray Binaries with INTEGRALand Future Directions}.
\newblock NewAR, 86, 101546.
\newblock \url{https://doi.org/10.1016/j.newar.2020.101546}.

\bibitem[{{Liu} et~al.(2006){Liu}, {van Paradijs} and {van den
  Heuvel}}]{2006liu}
{Liu}, Q.~Z., {van Paradijs}, J. and {van den Heuvel}, E.~P.~J. (2006)
  {Catalogue of high-mass X-ray binaries in the Galaxy (4th edition)}.
\newblock A\&A, 455(3), 1165--1168.
\newblock \url{https://doi.org/10.1051/0004-6361:20064987}.

\bibitem[{{Monageng} et~al.(2017){Monageng}, {McBride}, {Coe}, {Steele} and
  {Reig}}]{2017Monageng}
{Monageng}, I.~M., {McBride}, V.~A., {Coe}, M.~J., {Steele}, I.~A. and {Reig},
  P. (2017) {On the relationship between circumstellar disc size and X-ray
  outbursts in Be/X-ray binaries}.
\newblock MNRAS, 464(1), 572--585.
\newblock \url{https://doi.org/10.1093/mnras/stw2354}.

\bibitem[{{Neumann} et~al.(2023){Neumann}, {Avakyan}, {Doroshenko} and
  {Santangelo}}]{2023Neumann}
{Neumann}, M., {Avakyan}, A., {Doroshenko}, V. and {Santangelo}, A. (2023)
  {XRBcats: Galactic High Mass X-ray Binary Catalogue}.
\newblock A\&A, 677, A134.
\newblock \url{https://doi.org/10.1051/0004-6361/202245728}.

\bibitem[{{Okazaki} and {Negueruela}(2001)}]{2001OkazakiAndNegueruela}
{Okazaki}, A.~T. and {Negueruela}, I. (2001) {A natural explanation for
  periodic X-ray outbursts in Be/X-ray binaries}.
\newblock A\&A, 377, 161--174.
\newblock \url{https://doi.org/10.1051/0004-6361:20011083}.

\bibitem[{{Pecaut} and {Mamajek}(2013)}]{2013pecautmamajek}
{Pecaut}, M.~J. and {Mamajek}, E.~E. (2013) {Intrinsic Colors, Temperatures,
  and Bolometric Corrections of Pre-main-sequence Stars}.
\newblock ApJS, 208(1), 9.
\newblock \url{https://doi.org/10.1088/0067-0049/208/1/9}.

\bibitem[{{Rappaport} and {van den Heuvel}(1982)}]{1982Rappaport}
{Rappaport}, S. and {van den Heuvel}, E.~P.~J. (1982) {X-ray observations of Be
  stars}.
\newblock IAUS, 98, 327--344.

\bibitem[{{Reig}(2011)}]{2011Reig}
{Reig}, P. (2011) {Be/X-ray binaries}.
\newblock Ap\&SS, 332(1), 1--29.
\newblock \url{https://doi.org/10.1007/s10509-010-0575-8}.

\bibitem[{{Serino} et~al.(2022){Serino}, {Negoro}, {Nakajima}, {Kobayashi},
  {Asakura}, {Seino}, {Mihara}, {Tamagawa}, {Li}, {Matsuoka}, {Sakamoto},
  {Sugita}, {Komachi}, {Hiramatsu}, {Yoshida}, {Tsuboi}, {Iwakiri}, {Kawai},
  {Okamoto}, {Kitakoga}, {Kohara}, {Shidatsu}, {Iwasaki}, {Kawai}, {Niwano},
  {Hosokawa}, {Imai}, {Ito}, {Takamatsu}, {Nakahira}, {Ueno}, {Tomida},
  {Ishikawa}, {Tominaga}, {Nagatsuka}, {Kurihara}, {Ueda}, {Yamada}, {Ogawa},
  {Setoguchi}, {Yoshitake}, {Goto}, {Uematsu}, {Inaba}, {Tsunemi}, {Yamauchi},
  {Nonaka}, {Sato}, {Hatsuda}, {Fukuoka}, {Kawamuro}, {Yamaoka}, {Kawakubo} and
  {Sugizaki}}]{2022ATelserino}
{Serino}, M., {Negoro}, H., {Nakajima}, M., {Kobayashi}, K., {Asakura}, K.,
  {Seino}, K., {Mihara}, T., {Tamagawa}, T., {Li}, J., {Matsuoka}, M.,
  {Sakamoto}, T., {Sugita}, S., {Komachi}, K., {Hiramatsu}, H., {Yoshida}, A.,
  {Tsuboi}, Y., {Iwakiri}, W., {Kawai}, H., {Okamoto}, Y., {Kitakoga}, S.,
  {Kohara}, J., {Shidatsu}, M., {Iwasaki}, M., {Kawai}, N., {Niwano}, M.,
  {Hosokawa}, R., {Imai}, Y., {Ito}, N., {Takamatsu}, Y., {Nakahira}, S.,
  {Ueno}, S., {Tomida}, H., {Ishikawa}, M., {Tominaga}, M., {Nagatsuka}, T.,
  {Kurihara}, M., {Ueda}, Y., {Yamada}, S., {Ogawa}, S., {Setoguchi}, K.,
  {Yoshitake}, T., {Goto}, Y., {Uematsu}, R., {Inaba}, K., {Tsunemi}, H.,
  {Yamauchi}, M., {Nonaka}, Y., {Sato}, T., {Hatsuda}, R., {Fukuoka}, R.,
  {Kawamuro}, T., {Yamaoka}, K., {Kawakubo}, Y. and {Sugizaki}, M. (2022)
  {MAXI/GSC detection of an X-ray short transient event MAXI J0709-159}.
\newblock ATel, 15178, 1.

\bibitem[{{Virtanen} et~al.(2020){Virtanen}, {Gommers}, {Oliphant},
  {Haberland}, {Reddy}, {Cournapeau}, {Burovski}, {Peterson}, {Weckesser},
  {Bright}, {van der Walt}, {Brett}, {Wilson}, {Millman}, {Mayorov}, {Nelson},
  {Jones}, {Kern}, {Larson}, {Carey}, {Polat}, {Feng}, {Moore}, {VanderPlas},
  {Laxalde}, {Perktold}, {Cimrman}, {Henriksen}, {Quintero}, {Harris},
  {Archibald}, {Ribeiro}, {Pedregosa}, {van Mulbregt} and {SciPy 1. 0
  Contributors}}]{2020SciPy}
{Virtanen}, P., {Gommers}, R., {Oliphant}, T.~E., {Haberland}, M., {Reddy}, T.,
  {Cournapeau}, D., {Burovski}, E., {Peterson}, P., {Weckesser}, W., {Bright},
  J., {van der Walt}, S.~J., {Brett}, M., {Wilson}, J., {Millman}, K.~J.,
  {Mayorov}, N., {Nelson}, A. R.~J., {Jones}, E., {Kern}, R., {Larson}, E.,
  {Carey}, C.~J., {Polat}, {\.I}., {Feng}, Y., {Moore}, E.~W., {VanderPlas},
  J., {Laxalde}, D., {Perktold}, J., {Cimrman}, R., {Henriksen}, I.,
  {Quintero}, E.~A., {Harris}, C.~R., {Archibald}, A.~M., {Ribeiro}, A.~H.,
  {Pedregosa}, F., {van Mulbregt}, P. and {SciPy 1. 0 Contributors} (2020)
  {SciPy 1.0: fundamental algorithms for scientific computing in Python}.
\newblock NatMe, 17, 261--272.
\newblock \url{https://doi.org/10.1038/s41592-019-0686-2}.

\end{thebibliography}

\end{document}